\documentclass[prb,12pt]{revtex4}

\usepackage[english]{babel}
\usepackage[intlimits]{amsmath}
\usepackage[dvips]{graphicx}

\begin{document}

\begin{titlepage}

\title{Polymer Dynamics in Repton Model at Large Fields}
\author{Anatoly B. Kolomeisky}
\affiliation{Department of Chemistry, Rice University, Houston, 
TX 77005-1892, USA}
\author{Andrzej Drzewi\'nski}
\affiliation{Czestochowa University of Technology, Institute of
Mathematics and Computer Science,
ul.Dabrowskiego 73, 42-200 Czestochowa, Poland\\}
\affiliation{Institute of Low Temperature and Structure Research, 
Polish Academy of Sciences, P.O. Box 1410, 50-950 Wroc\l aw 2, Poland}

\begin{abstract} 
Polymer dynamics at large fields in Rubinstein-Duke repton model is 
investigated theoretically. Simple diagrammatic approach and analogy with 
asymmetric simple exclusion models are used to analyze the reptation 
dynamics of polymers. It is found that  for  polyelectrolytes  the drift velocity decreases exponentially as a function 
of the external field with an exponent depending on polymer size and parity, while for polyampholytes the drift velocity is independent of polymer chain size. 
However, for polymers, consisting of charged and neutral blocks, the drift 
velocity approaches the constant limit which is determined by the size of 
the neutral block. The theoretical arguments are supported by extensive 
numerical calculations by means of density-matrix renormalization group 
techniques.

\end{abstract}

\maketitle

\end{titlepage}

\section{Introduction}

Dynamics of polymers in  a  dense medium is important in many chemical, 
biological and industrial processes. Current theoretical understanding of 
these phenomena is based on de Gennes reptation idea.\cite{degennes} According 
to this idea,  the polymer chain cannot move easily in  directions normal to 
its length because of the many  obstructions in the system, instead, the 
polymer molecule diffuses in a snake-like motion along its contour. 

The simplest model of polymer dynamics in systems with obstacles, which 
incorporates de Gennes reptation mechanisms, is  the so-called  repton model. 
It is a lattice model and it was first proposed by Rubinstein 
\cite{rubinstein87} and later adapted by Duke as a model for the gel 
electrophoresis of DNA.\cite{duke} Electrophoresis is a method of 
size-separation of charged polymers using an electric field.\cite{viovy00} Numerous efforts 
have been invested in obtaining exact steady-state solutions of the repton 
model,\cite{widom91,vanleeuwen91,vanleeuwen92,kooiman93,vanleeuwen93,prahofer96,
kolomeisky96,willmann02}  however, with limited success. Van Leeuwen and 
Kooiman \cite{vanleeuwen92,kooiman93,vanleeuwen93} made an important advance 
in the analysis of the model by applying periodic boundary conditions. For 
the repton model with open boundaries formally exact but implicit formulas 
for the diffusion coefficient and the drift velocity are known.\cite{widom91}  
However, except for small chains,\cite{widom91} there are no exact solutions 
for the stationary state. The more successful approach to understand the 
polymer dynamics in the repton model has been utilized in  Monte Carlo computer 
simulations studies, \cite{duke,barkema94,newman97} although the results for 
large polymer sizes $N$ and/or large external fields are difficult to obtain. 
Recently, a new numerical method of investigation of the repton model based on 
the density matrix renormalization group  (DMRG) approach has been 
presented.\cite{carlon01,carlon02} 

In this paper we investigate the dynamics of polymers in  the  repton model in 
the limit of very large external fields. In the repton model, this limit is 
less physical because it ignores two features of real electrophoresis that 
are important at large fields (although not at low fields): 1) transmission 
of tension through the segments of the polymer; and 2) the appearance of 
hernias, i.e., chain branching and creation of loops. Despite these facts 
the repton model in the limit of large fields still  captures  many qualitative 
properties of the gel electrophoresis of biological macromolecules. Exact 
analytic and asymptotic behaviors for the drift velocities are found using 
simple diagrammatic  approach and analogy with asymmetric simple exclusion 
processes.\cite{barkema96,schutz} This paper is organized as follows: the 
repton model is introduced in Sec. II, while in Sec. III the DMRG method is 
described. In Sec. IV the polymer dynamics at large fields is investigated 
using diagrammatic approach and analogy with asymmetric simple exclusion 
processes. Sec. V collects a series of numerical results of  reptation  in
a strong field, while Sec. VI summarizes all results and concludes our paper.

\section{Repton Model}

The repton model is tailored to describe the gel electrophoresis of DNA, and 
the gel is thought of as a space divided into cells, so that each cell 
corresponds to a pore in the gel. The macromolecule is divided into $N$ 
segments of equal length, and each segment is replaced by its midpoint 
(see Fig 1a). These points are called {\it reptons}, and each repton can 
consist of many monomer units. The number of reptons that each cell may 
accommodate is unlimited and self-avoidance effects are neglected.

As shown in Fig. 1b,  the  polymer chain consisting of $N$ reptons is connected 
by $N-1$ bonds. Each element of the chain is permanently confined to its own 
track running in the $x$ direction (parallel to the applied electric field), 
where $x$ is a coordinate that takes on only the discrete values. The polymer 
connectivity requires that the coordinates of adjacent reptons differ by only 
$0$ or $\pm1$. Neighboring reptons that have a common value of their $x$ 
coordinate represent a cluster of successive reptons occupying a common cell 
in the gel. The clusters indicate an excess of reptons in cells; a polymer 
moves by the diffusion of these extra reptons. An interior repton $i$ 
($2 \leq i \leq N-1$) can move only if it is the end repton of a cluster, 
and then it is allowed to move only in the direction of that one of its two 
neighbors which is not part of the cluster. This is shown in Fig. 1b by the 
arrows, $\uparrow$ (up) or $\downarrow$ (down). An end repton ($i=1$ or $N$) 
is always allowed to move. If it is not part of a cluster, it can only move 
in the direction of its neighbor. If the end repton is part of a cluster, it 
can move in either direction.

For every allowed move $\uparrow$ or $\downarrow$ of a repton there is 
associated the transition probability per unit time, $B$ or $B^{-1}$, 
respectively, and $B$ defined as $B=exp(\varepsilon/2)$, where $\varepsilon$ 
is the dimensionless 
constant electric field. Since for every bond connecting two reptons at sites 
$i$ and $i+1$ ($i=1,\cdots,N-1$) the difference $x_{i+1}-x_{i}$ can have any 
of three values $0$ or $\pm1$, the repton chain has $3^{N-1}$ possible 
distinct configurations irrespective of its location as a whole in the $x$ 
direction. Let $y=1, \cdots, 3^{N-1}$  numerate these polymer configurations.

There is an equivalent representation of the repton model that maps it onto 
a one-dimensional asymmetric simple exclusion model with two types of 
particles.\cite{barkema96,schutz} In this representation each link between 
adjacent reptons in our polymer chain  corresponds  to a site on the new lattice. 
We associate a positive (``+'') particle, or a negative (``-'') particle, or a hole (``0'') with each site of the new lattice depending on the slope 
($x_{i+1}-x_{i}$) of the link in the original repton chain. This mapping is 
illustrated in Fig. 2 for one particular configuration of the polymer.

The dynamical rules for the repton model are easily expressed in the language 
of the asymmetric exclusion model. In the bulk the system evolves according 
to the following rules:\\
$(+)_{i}(0)_{i+1} \rightarrow (0)_{i}(+)_{i+1}$ with rate $B$;\\
$(0)_{i}(+)_{i+1} \rightarrow (+)_{i}(0)_{i+1}$ with rate $B^{-1}$;\\
$(0)_{i}(-)_{i+1} \rightarrow (-)_{i}(+)_{i+1}$ with rate $B$;\\
$(-)_{i}(0)_{i+1} \rightarrow (0)_{i}(-)_{i+1}$ with rate $B^{-1}$.\\
At the left end of the lattice ($i=1$):\\
$(0)_{1} \rightarrow (+)_{1}$ with rate $B$;\\
$(+)_{1} \rightarrow (0)_{1}$ with rate $B^{-1}$;\\
$(-)_{1} \rightarrow (0)_{1}$ with rate $B$;\\
$(0)_{1} \rightarrow (-)_{1}$ with rate $B^{-1}$.\\
At the right end of the lattice ($i=N-1$):\\
$(0)_{N-1} \rightarrow (-)_{N-1}$ with rate $B$;\\
$(-)_{N-1} \rightarrow (0)_{N-1}$ with rate $B^{-1}$;\\
$(+)_{N-1} \rightarrow (0)_{N-1}$ with rate $B$;\\
$(0)_{N-1} \rightarrow (+)_{N-1}$ with rate $B^{-1}$.\\
Note, that here the exchange of positive and negative particles is not allowed.

\section{Density Matrix Renormalization}

In order to determine the properties of the reptation models we have used 
the density-matrix renormalization group (DMRG). This technique enables us to 
study some global characteristics of polymers as the drift velocity and  
local ones as average shapes. The method was introduced in 1992 by  S. White 
as an efficient algorithm to deal with a quantum Hamiltonian for one 
dimensional spin systems.\cite{white} 

This iterative basis--truncation method allows us to approximate eigenvalues 
and eigenstates with very high accuracy and in a controlled way.\cite{dmrg} 
The method is not restricted to quantum systems only; it has also been 
successfully applied to a series of problems, ranging from two-dimensional 
classical systems\cite{clsys} to stochastic processes.\cite{stproc,carlon01,
carlon02} Using the formal similarity between the Master equation and
the one-dimensional Schr\"odinger equation, the method of DMRG can also be 
applied to the Master Equation for the reptating chains\cite{carlon01,carlon02}.

The number of configurations in the Hilbert space of the polymer grows very 
fast with its number of bonds $L=N-1$ (as $3^L$). Therefore, it is 
prohibitively  difficult 
to solve exactly chains longer than $L=12$. But it is reasonable to 
eliminate the least probable (in the density matrix sense) states and keep 
only the most important ones. Although from this stage our calculation is not 
exact anymore, we can obtain very efficient approach if the weights of 
the discarded states are very small.

Starting with a small system (e.g. $L=4$ in our case), for which $H$ can be 
diagonalized exactly, one adds iteratively couples of reptons until the 
allowed (in the computational sense) size of the effective matrices is 
reached. Then further addition of new reptons forces one to discard 
simultaneously the least important states to keep the size of the effective 
matrices fixed. This truncation is done through the construction of a reduced 
density matrix whose eigenstates provide the optimal basis set (see Refs. 
\cite{white,dmrg} for details). The size of the effective matrix is then 
substantially smaller than the original dimensionality of the configurational 
space $(3m)^2 \ll 3^L$. Generally, the larger is $m$, the better accuracy is 
guaranteed. In the present case, we keep this parameter up to $m=81$ (the 
stronger the fields, the larger the $m$ is necessary). 

It is worth stressing that for a reptating chain, one has to face with the 
non--hermitian matrix $H$ (which becomes hermitian for zero driving field only). 
Since we are studying the very strong-field case we have to apply the 
non--hermitian variant of the standard DMRG algorithm,\cite{stproc}  where one 
has to distinguish between the right and left eigenvector belonging to the 
same eigenvalue. Since $H$ is a  stochastic matrix, its lowest eigenvalue is 
equal to $0$ and the corresponding left eigenvector is proportional to the 
unit vector. The right eigenvector gives the stationary probability distribution. 
Generally, the DMRG method works best when the eigenvalues of $H$ are well 
separated. For long chains and strong fields the spectrum  of $H$ gets an 
accumulation of eigenvalues near the zero eigenvalue of the stationary state. 
This hampers the convergence of the method seriously. Usually, enlarging the 
basis $m$ improves the accuracy substantially. However, for the reptation 
problem it helps very little and limits the systems under study with respect 
to the length. 

To construct the reduced density matrix from the lowest 
eigenstates one has to diagonalize the effective matrix $H$ at each DMRG 
step. Therefore, we used the so-called Arnoldi method,\cite{arnoldi}  which 
is known to be particularly stable for non-hermitian problems.

\section{High-Field Polymer Dynamics}

\subsection{Diagrammatic Approach}

Consider a polyelectrolyte (PE) molecule, where all charges in the polymer are 
of equal sign and of equal value,  in the limit of very large applied electric 
fields. The formal expression for the drift velocity $V$ of a chain of size 
$N$ is given by\cite{widom91}
\begin{equation}\label{drift_velocity}
V=N^{-1} \sum_{y} (Br_{y}-B^{-1}s_{y})p_{y},
\end{equation}
where $r_{y}$($s_{y}$) is the total number of $\uparrow$ ($\downarrow$) arrows 
on all of the reptons of chain in configuration $y$ and $p_{y}$ is the 
probability to find the chain in the configuration $y$. This equation implies 
that every configuration contributes to the drift velocity. Except for the case 
of small $N$, no general closed expression for $V$ is known.\cite{widom91} At 
the limit of large fields ($ \varepsilon \rightarrow \infty$) we expect that only 
few configurations play a major role in the polymer chain motions.

To single out relevant configurations, let us present them diagrammatically as 
follows. Put each configuration in a box, and connect the boxes with allowed 
transitions between the configurations by arrows. The direction along an
arrow corresponds to a more probable transition (with rate $B$), while the 
opposite direction means a less probable transition (with rate $B^{-1}$). 
The illustration of the diagrammatic representation for the polymer chain of 
size $N=3$ is given in Fig. 3.

Diagrams for general $N$ are more complicated but have the features that are 
already present for the $N=3$ diagram. There are cycles in the middle of the 
picture (for example, $y=2 \rightarrow y=1 \rightarrow y=4 \rightarrow y=2$). 
Each cycle has a size $N$. There are configurations which are less probable at 
any field $ \varepsilon \neq 0$, such as the configuration $y=3$. There are 
configurations which are more probable (configuration $y=7$ is an example).

In the limit of high fields there is a special class of so-called trap 
configurations. We define a trap configuration as one in which the allowed 
moves of the reptons are all highly unfavorable ones (against the field). The 
trap configurations for small chains are shown in Fig. 4.

The number of trap configurations quickly increases with $N$. As we can see 
from Fig. 4,  there is a one trap configuration for $N=3$, two trap 
configurations for $N=4$, and four trap configurations for $N=5$. For 
$\varepsilon \gg 1$ we would expect to find a polymer chain mainly in these trap 
configurations in the stationary state. But a closer look at our diagrammatic 
picture shows that only one (for odd $N$) or two (for even $N$) trap 
configurations will be real traps. These real traps have a symmetric ``U'' 
shape (consistent with experimental observations of large-field  behavior of 
DNA molecules in gel electrophoresis\cite{smith89}).

To demonstrate this we look at two possible trap configurations of an 
$N=5$ polymer chain (see Fig. 5). Suppose we find a polymer molecule in the 
configuration $y=2$ (see Fig. 5a). For $\varepsilon \rightarrow \infty$ there is a 
very small but nonzero probability for a transition from the trap 
configuration $y=2$ to the configuration $y=1$. Then with only probability 
1/2 the chain returns to the trap. Thus at longer times the probability to 
find the chain in this trap configuration decreases to zero, therefore 
implying that the chain escapes this trap. In contrast, if the chain is in 
the trap configuration $y=5$ (see Fig. 5b), and it undergoes a transition to 
the configuration $y=4$ or $y=6$, then with overwhelming probability the 
system returns back to the real trap configuration. This kind of argument can 
be extended to a chain of any size.

Using this idea we can calculate the probabilities of different configurations 
in the high-field limit. In the example of $N=5$ (Fig. 5b) the probability to 
find the trap configuration $y=5$ is $p_{5} \sim 1$. Then because of the 
local detailed balance $p_{4}=p_{6} \sim B^{-2}$. For general $N$ this trend 
will continue: each move up on the diagram away from the real trap decreases 
the probability of configurations by $B^{2}$. In this case the expression for 
the drift velocity becomes:
\begin{equation} 
V=N^{-1} \left[-p_{5} (2/B) + p_{4}(B-3/B) + p_{6} (B-3/B)+ \cdots)=
N^{-1}(-6/B^{3}+ \cdots\right].
\end{equation}
The structure of the equation implies that the contributions to the drift 
velocity from the trap configurations and from the configurations leading 
to the traps cancel each other. The process of cancellation continues as we 
go up on the diagram until the branching. By branching we call the existence 
of another probable way of moving the chain, not leading to the real trap 
configuration. For example, in  Fig. 4b the configuration $y=1$ is a place 
of branching because there are two escape routes from this configuration. By 
following back from the trap configuration on the diagram, we effectively 
move a zero-slope bond from the ends to the middle of the chain (see Figs 3 
and 5b). Then for odd $N$ there  are  $(N-1)/2$ steps from the trap configuration 
before the branching, while for even $N$ there are $(N-2)/2$ steps. Thus the 
probability of the configuration which gives a non-zero contribution to the 
drift velocity is $\sim B^{1-N}$ (for odd $N$) or $\sim B^{2-N}$ (for 
even $N$). As a result, in the limit of large electric fields the drift 
velocity is given by
\begin{eqnarray}
V \sim B^{2-N}, & N \mbox{ -- odd} \nonumber \\
V \sim B^{3-N}, & N \mbox{ -- even}.
\end{eqnarray}
These expressions are valid for {\it any} $N$ and large fields (when $NE \gg 1$). 
However, we cannot determine explicitly the constants in front of field-dependent 
terms. We can only conjecture that it grows also  exponentially with the polymer 
size.  Exponential decrease in the drift velocity of polymers has been observed 
in recent  computer simulations of gel electrophoresis with cage 
model.\cite{vanheukelum00}

The diagrammatic approach can also be used to treat polyampholytes (PA), the 
polymer molecules with charges of different sign. Specifically, consider 
alternating PA, where positive and negative charges along the polymer 
backbone alternate with each other. Then in the limit of large fields we 
expect only one ``crown''-shaped trap configuration as a real trap.

To calculate the drift velocity, we again utilize Eq. (\ref{drift_velocity}), 
which is also valid for polyampholytes, but the meaning of the parameters 
$r_{y}$ and $s_{y}$ changes. Now for a given configuration $y$, $r_{y}$ is 
the difference between the number of $\uparrow$ moves (positive direction of 
the motion) with the probability $B$ minus the number of $\downarrow$ moves 
with the probability $B$. Similarly, $s_{y}$ is the difference between the 
number of  $\downarrow$ moves with the probability $B^{-1}$ minus the number 
of $\uparrow$ moves with the probability $B^{-1}$. It is obvious that $r_{y}$ 
and $s_{y}$ can be negative, positive or zero, in contrast to the PE case 
where these parameters were always non-negative. 

For even $N$ the drift velocity $V=0$ at {\it all} values of the applied 
electric field. This can be seem from the fact that for any configuration  
$y$ with parameters $r_{y}$ and $s_{y}$ there is a configuration $y^{*}$ 
with the parameters $-r_{y}$ and $-s_{y}$ related to $y$ by inversion. If 
the configuration $y^{*}$ and $y$ are the same then $r_{y}=s_{y}=0$.

The case of odd $N$ is different. For $\varepsilon \rightarrow \infty$ the diagrammatic 
approach predicts that the drift velocity is asymptotically given by 
\begin{equation}\label{drift_velocity_PA}
V \sim B^{-3}
\end{equation}
for {\it all} $N$. For the smallest alternating PA chain, $N=3$, we can 
calculate the drift velocity at fields directly by using transition-rate 
matrix method.\cite{widom91} It gives us 
\begin{equation}\label{tm3}
V= \frac{4(B-B^{-1})}{3(B^{4}+B^{-4}) + 7(B^{2}+B^{-2}) + 16},
\end{equation}
which in the limit of very large $\varepsilon$ reduces to Eq. (\ref{drift_velocity_PA}).

\subsection{Analogy with Asymmetric Exclusion Processes}

Similarly to the case of polyelectrolytes and polyampholytes, the repton model 
can be applied to polymers with charged and neutral segments. Here we consider 
a chain comprised of two blocks. The left block consists of $k$ charged reptons 
(the charges of the same sign), and the right block consists of $N-k$ neutral 
reptons. The experimental realization of such system, for example, is gel 
electrophoresis of DNA-neutral protein complex.\cite{ulanovsky90}

Using the mapping onto the one-dimensional asymmetric simple exclusion model 
as described in Sec. II we identify a negative bond variable 
($j_{i} \equiv x_{i+1}-x_{i}=-1$), a positive bond variable ($j_{i}=+1$), and 
a zero bond variable ($j_{i}=0$) with a $p$-particle, $m$-particle, and vacancy, 
respectively. In the steady-state the current of $p$-particles $J^{(p)}$ is the 
same at every site of the lattice but different from the current of 
$m$-particles $J^{(m)}$ as can be seen from the asymmetry of the system. We 
assume that the positive direction for the current of $p$-particles  is from 
left to right, and the positive direction for the current of $m$-particles is 
from right to left. Then the drift velocity of the original repton model is 
connected to the steady-state currents in this asymmetric simple exclusion 
model by
\begin{equation}\label{vel.current}
V=J^{(p)}+J^{(m)}.
\end{equation}
It is convenient to introduce a difference of the currents $\Delta J$
\begin{equation}\label{dif.current}
\Delta J=J^{(p)}-J^{(m)}.
\end{equation}

The average densities $<p_{i}>$ and $<m_{i}>$ determine the currents. At the 
left end of the polymer  ($i=1$) the currents are 
\begin{eqnarray}\label{currents.1}
J^{(p)}&=&B <1-p_{1} - m_{1}> -B^{-1}<p_{1}> \nonumber \\
J^{(m)}&=&B <m_{1}> -B^{-1}<1-p_{1}-m_{1}>.
\end{eqnarray}
For the sites $i=1,\cdots,k-1$ of the charged block we have
\begin{eqnarray}\label{currents.charged}
J^{(p)}&=&B <p_{i}(1-p_{i+1} - m_{i+1})> -B^{-1}<(1-p_{i} - 
m_{i})p_{i+1}> \nonumber \\
J^{(m)}&=&B <(1-p_{i} - m_{i})m_{i+1}> -B^{-1}<m_{i}(1-p_{i+1} - m_{i+1})>.
\end{eqnarray}
For the sites $i=k,\cdots,N-2$ of the neutral block we have 
\begin{eqnarray}\label{currents.neutral}
J^{(p)}&=& <p_{i}(1-p_{i+1} - m_{i+1})> -<(1-p_{i} - 
m_{i})p_{i+1}> \nonumber \\
J^{(m)}&=& <(1-p_{i} - m_{i})m_{i+1}> -<m_{i}(1-p_{i+1} - m_{i+1})>.
\end{eqnarray}
The currents at the right end of the lattice ($i=N-1$) are given by
\begin{eqnarray}\label{currents.N-1}
J^{(p)}&=&<p_{N-1}> - <1-p_{N-1}-m_{N-1}> \nonumber \\
J^{(m)}&=&<1-p_{N-1}-m_{N-1}> -<m_{N-1}>.
\end{eqnarray}

Introducing an average density of vacancies $<n_{i}>=<1-p_{i}-m_{i}>$ we 
obtain from the Eqs. (\ref{currents.1})
\begin{equation}\label{vel.1}
V=\frac{B-B^{-1}}{2}(1+<n_{1}>)-\frac{B+B^{-1}}{2}(<p_{1}>-<m_{1}>),
\end{equation}
and
\begin{equation}\label{dif.1}
\Delta J = \frac{B+B^{-1}}{2}(3<n_{1}>-1)+\frac{B-B^{-1}}{2}(<p_{1}>-<m_{1}>).
\end{equation}
From Eqs. (\ref{currents.neutral},\ref{currents.N-1}) one can conclude that
\begin{eqnarray}\label{dif.currents}
\Delta J = <n_{i+1}>-<n_{i}>,&  &i=k,\cdots,N-2 \nonumber \\
\Delta J = 1- 3<n_{N-1}>. &  &
\end{eqnarray}
Consequently,
\begin{equation}\label{holes}
<n_{i}>=(3N-3i-2) <n_{N-1}>-(N-i-1), \quad i=k,\cdots,N-1.
\end{equation}

It is impossible to find exact solutions of the system 
(\ref{vel.current}-\ref{holes}) for general values of the field because the 
number of variables exceeds the number of equations, except in the limit 
$\varepsilon \rightarrow \infty$. In this limit one realizes that 
$<n_{1}>=<n_{2}>=<n_{3}>= \cdots = <n_{k}>=0$. Then from Eq. (\ref{holes}) 
one can obtain
\begin{equation}\label{holes.N-1}
<n_{N-1}>=\frac{N-k-1}{3(N-k)-2},
\end{equation}
\begin{equation}
\Delta J= \frac{1}{3(N-k)-2}.
\end{equation}

Now comparing Eqs. (\ref{dif.1}) and (\ref{vel.1}) in the limit of large 
applied field we obtain the following expression for the drift velocity
\begin{equation}\label{velocity.final} 
V=\Delta J=\frac{1}{3(N-k)-2}.
\end{equation}

For $k=1$ this system was investigated in Ref\cite{barkema96}. In that paper 
Monte Carlo simulations were used to extract the drift velocity at any value 
of the field. Also an exact solution in the infinite field-limit was found, 
which for $k=1$ reproduces Eq. (\ref{velocity.final}).

Another way to check our results is to calculate directly the drift velocity 
for small chain $N=3$ by using transition-rate matrix method.\cite{widom91} 
Then calculations for $k=1$ yield
\begin{equation}
V=\frac{B^{2}-B^{-2}}{4(B^{2}+B^{-2})+3(B+B^{-1})+4},
\end{equation}
which in the limit $\varepsilon \gg 1$ approaches to 1/4, in agreement with 
Eq. (\ref{velocity.final}).

For $k=2$ and $N=3$ the drift velocity is given by
\begin{equation}
V=\frac{B^{2}-B^{-2}}{(B^{2}+B^{-2})+3(B+B^{-1})+1}.
\end{equation}
In the limit $\varepsilon \rightarrow \infty$ we have $V \rightarrow 1$, again in 
agreement with our general expression (\ref{velocity.final}).

\section{NUMERICAL RESULTS}

Since we are interested in behavior of polymers in a very strong electric
field we have to analyze a strongly non-hermitian Hamiltonian. Although, from
computational reasons, it limits our considerations to chains of a
moderate length, studying them we can obtain a clear picture.

Note that the number of reptons is always odd in our DMRG calculations. This provides an additional constraint to our numerical method.

\subsection{Polyelectrolytes}

In order to verify which configurations dominate for strong fields, we can
calculate the average slopes $\langle y_i \rangle$ as function of the bond
position along a chain. They are always symmetrical, which is a consequence of  
the intrinsic symmetry between
a head and a tail for the uniformly charged chains. Summing elements 
$\langle y_j \rangle$ up to a certain repton the average shape of a polymer can be 
found. As a point of reference,  the position of the first repton is established as a zero.

As shown in Fig.6, with the increasing  field $\varepsilon$ the polymer configuration 
is changing from the horizontal to the U-shape  form. In the left part of a chain the 
average slopes $\langle y_i \rangle$ are going to the value $-1$, whereas in the right 
part they reach the value $+1$. That is why the configuration  minimum goes to the 
value of $(N-1)/2$. At the same time, it means that when $\varepsilon$ grows, the real 
trap configuration is  dominating  more and more, in a perfect agreement with our 
expectation.

In Fig.7 we have collected the DMRG results for the drift velocity. They
are calculated from the expression given by van Leeuwen and 
Kooiman.\cite{thesis} Its dependence on the electric field 
(in the inset) is in agreement with previous results.\cite{barkema94} For small 
fields one can observe a linear dependence, which is followed by  a maximum and an 
exponential decay for large fields. The presence of an exponential decay
results from the dominating role of the trap configurations. Polymers are stuck on 
obstacles being pulled at both arms. Since the tension is not translated in the repton 
model, the resulting drift velocity of the polymers decreases strongly with the field.

In order to find out the decay exponent we have presented our results in
the linear-log scale (Fig. 7). As one can see, in the large field limit the agreement
with the expression found by the diagrammatic approach (Eq.(3)) is perfect for all 
calculated polymer chains.

\subsection{Polyampholytes}

DMRG method can also be extended to calculate properties of polyampholytes. As an 
example, we have studied alternating PA chains. Since only odd-number reptons 
polymer chains are considered the sign of end charges is always the same 
(positive in our case).

The average shapes for polymers of size $N=11$ are presented in Fig. 8.
As expected,  a polymer chain configuration is approaching the crown shape at
large fields. Since in our plot the  chain is relatively small, the  influence of 
end reptons is still relatively large. Generally, when $\varepsilon$ grows, one can 
expect that for all $N$ the amplitude of the middle part goes to values of a range 
of a few bond lengths.

Our calculations for the drift velocity, in contrast to uniformly charged chains, 
indicate that the decay exponent does not depend on an alternating chain length
(Fig. 9). This conclusion is  in  a full agreement with our theoretical predictions: 
see Eq.(\ref{drift_velocity_PA}).

\subsection{Polymers Consisting of Neutral and Charged Blocks}

DMRG also allows us to investigate the polymer chains consisting of neutral and 
charged blocks. Let us first consider a chain consisting of all neutral reptons except 
one charged repton at the end. Hence, according to our notation from Sec.IV, $k=1$. 
Obviously, the average shape here differs substantially from the case where  all reptons 
charged. Nevertheless, the resulting shapes are intuitively easy to understand. As one 
can see in Fig. 10, the charged repton at the left end pulls the whole chain, and this 
effect is stronger at larger fields.  

In order to  determine  a  size-dependence of drift velocity  for different  polymers
with  neutral and charged blocks, we have studied two cases: $k=1,2$. The results are 
presented in Figs. 11 and 12, and they indicate  that the limit behavior  at large 
fields is in a perfect agreement with Eq.(\ref{velocity.final}). Note also that 
the absolute values of drift velocities are larger for the polymers with more charged 
reptons, i.e., the same-size polymer chains with $k=2$ move significantly faster than 
the polymers with only one charged end repton, in agreement with intuitive expectations. 
Our results  agree also  with Monte Carlo simulations\cite{barkema96} and numerical 
calculations\cite{drzewinski03} for a magnetophoresis, which corresponds to our $k=1$ case.

\section{Summary and Conclusions}

We have analyzed a lattice model of reptation (the Rubinstein-Duke model)
to study the dynamics of polymer in a dense medium. Asymptotically exact 
results in the limit of large applied fields have been obtained by means of 
simple diagrammatic approach and using the analogy with asymmetric simple exclusion 
processes. Our theoretical arguments are based on a realistic assumption that at 
large fields only a few configurations are important for polymer dynamics.
 
The method was successfully used for different types of polymers. For polyelectrolytes 
and polyampholytes, we find that the drift velocity is exponentially decreasing as 
a function of an external field. For PE chains the exponent is a function of polymer 
size, while for PA it is independent of polymer size. For polymers consisting of 
neutral and charged blocks  the situation is very different. In this case, at large 
fields the drift velocity approaches a constant value, which  depends on the size of 
the neutral block.  

Our theoretical predictions are well supported by extensive numerical calculations by
density-matrix renormalization techniques. First, we have determined the average shapes 
of polymer molecules. Our results indicate that at large fields the polymer chains can 
be found mainly in a few trap configurations, in excellent agreement with our 
predictions. Furthermore, the dependence of the drift velocity on external field 
has been investigated. It confirms that the drift velocity of polyelectrolytes and 
polyampholytes decays exponentially in the large field limit,  and for polyelectrolytes 
the decay exponent depends on a polymer length. In contrast to fully charged polymers, 
for chains consisting of neutral and charged blocks  the drift velocity approaches the 
constant values determined by  the size of the neutral blocks.

The importance of our analytical and numerical results on the behavior of repton model at large fields for real systems, e.g., gel electrophoresis, is unclear.\cite{viovy00} The repton model does not take into account the creation of hernias and the transfer of tension forces, which are very important for the dynamics of polyelectrolytes at large fields. However, one may naively expect that the creation of hernias is less important for PA and for the polymers with neutral and charged blocks. Despite its limited applicability to gel electrophoresis, our calculations provide exact asymptotic results for several classes of asymmetric simple exclusion processes with two types of of oppositely moving particles,\cite{schutz03} where the number of exact results is very limited.

\section*{Acknowledgments}

ABK acknowledges the Donors of the American Chemical Society Petroleum 
Research Fund (Grant No. 37867-G6) for support of this research. ABK also 
acknowledges  the financial support of the Camille and Henry Dreyfus New 
Faculty Awards Program (under Grant NF-00-056), the Welch Foundation (under 
grant No. C-1559) and Center for Biological and Environmental Nanotechnology 
at Rice University. This work has been also
supported by the Polish Science Committee (KBN) under grant in years 2003-2005. 
We also thank M.E. Fisher and  B. Widom for 
critical discussions,  and J.M.J. van Leeuwen for careful reading of the manuscript.

\vspace{5mm}

\begin{figure}[h]
\begin{center}
\vskip 1.5in
\unitlength 1in
\begin{picture}(3.4,2.8)
\resizebox{4.0in}{4.0in}{\includegraphics{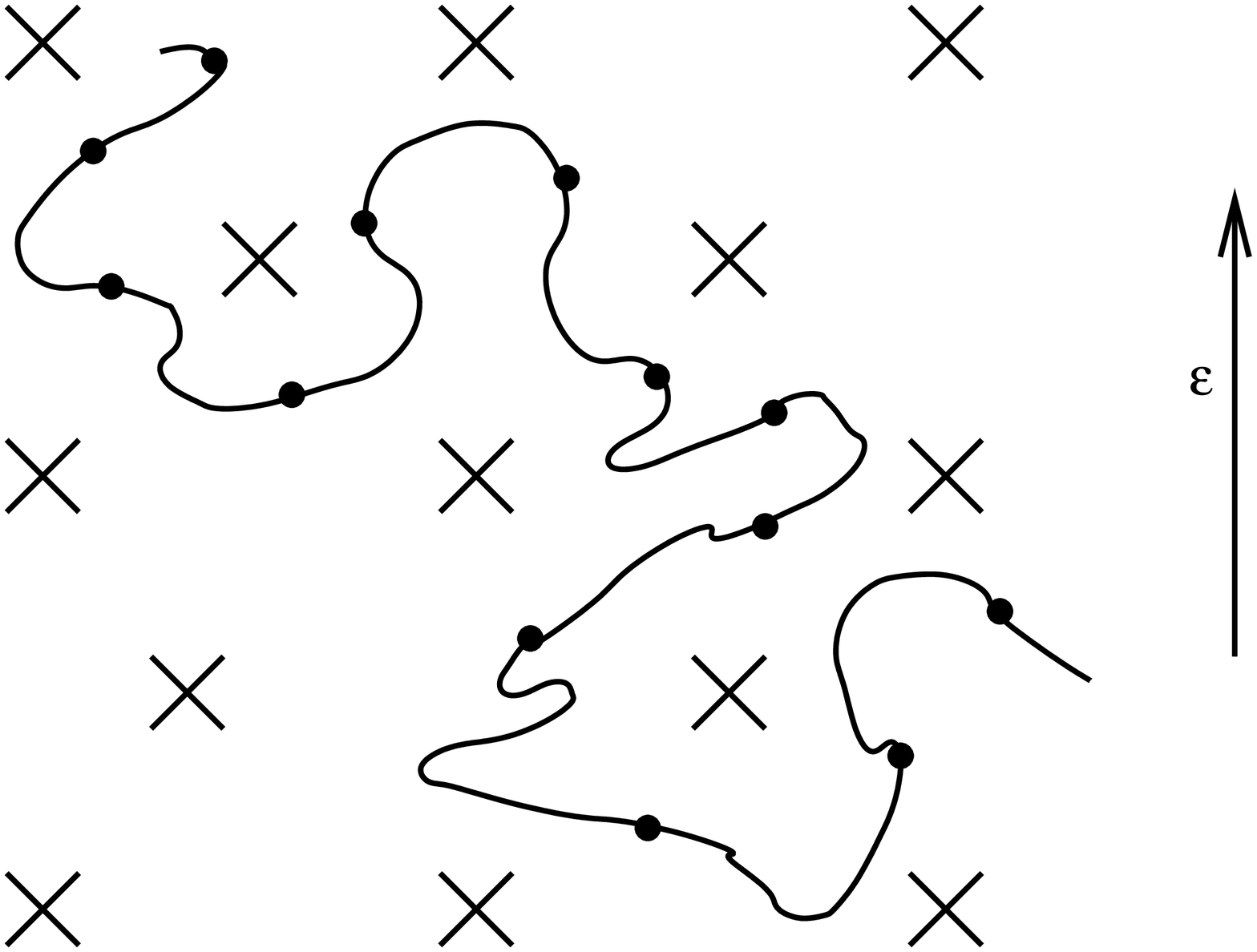}}
\end{picture}
\vskip 3in
 \begin{Large} Fig.1a \end{Large}
\end{center}
\vskip 3in
\end{figure}

\begin{figure}[h]
\begin{center}
\vskip 1.5in
\unitlength 1in
\begin{picture}(4.5,3.5)
\resizebox{5.0in}{4.5in}{\includegraphics{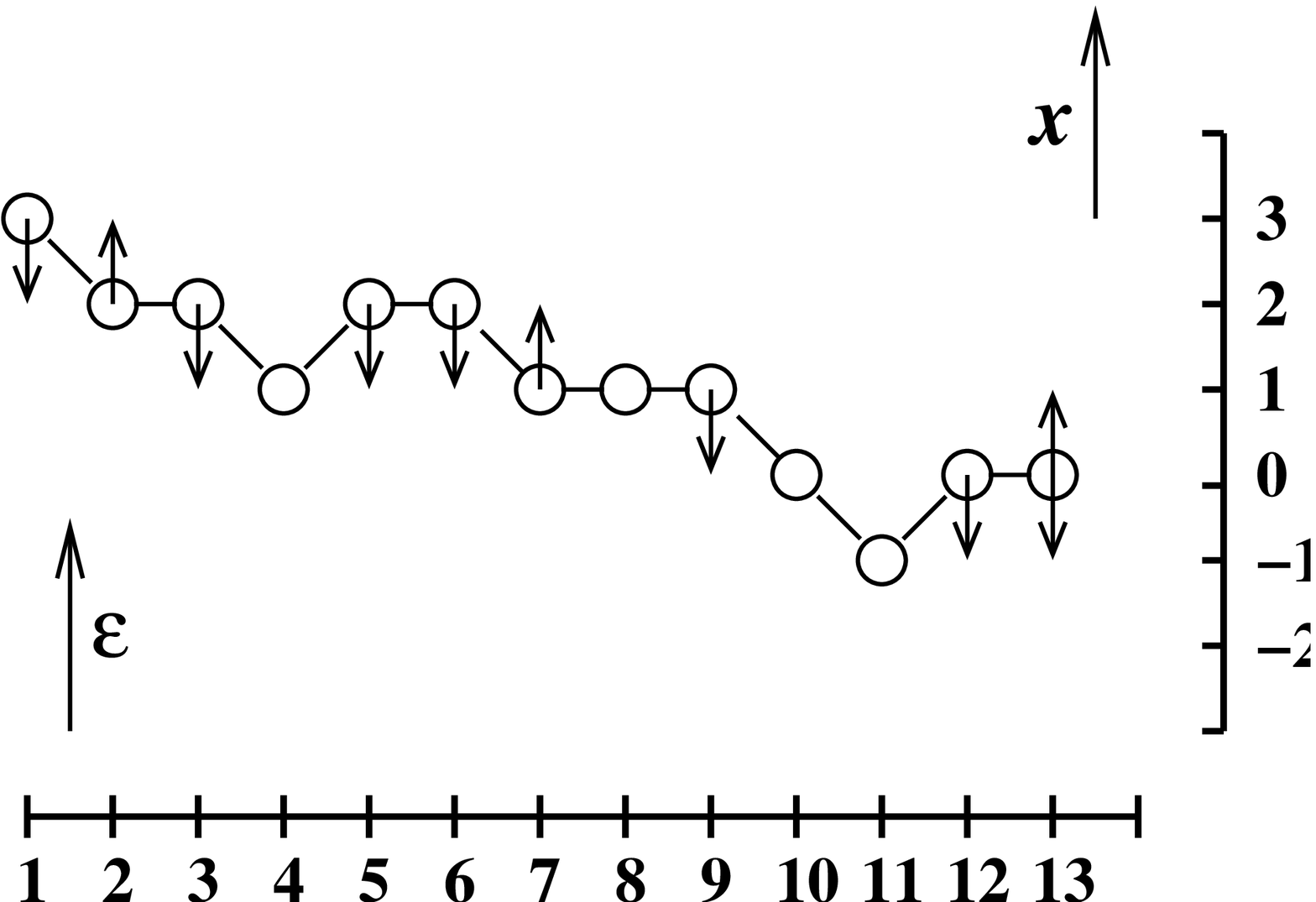}}
\end{picture}
\vskip 3in
 \begin{Large} Fig.1b \end{Large}
\end{center}
\vskip 3in
\end{figure}

\begin{figure}[h]
\begin{center}
\vskip 1.5in
\unitlength 1in
\begin{picture}(5.1,4.2)
\resizebox{5.8in}{5.2in}{\includegraphics{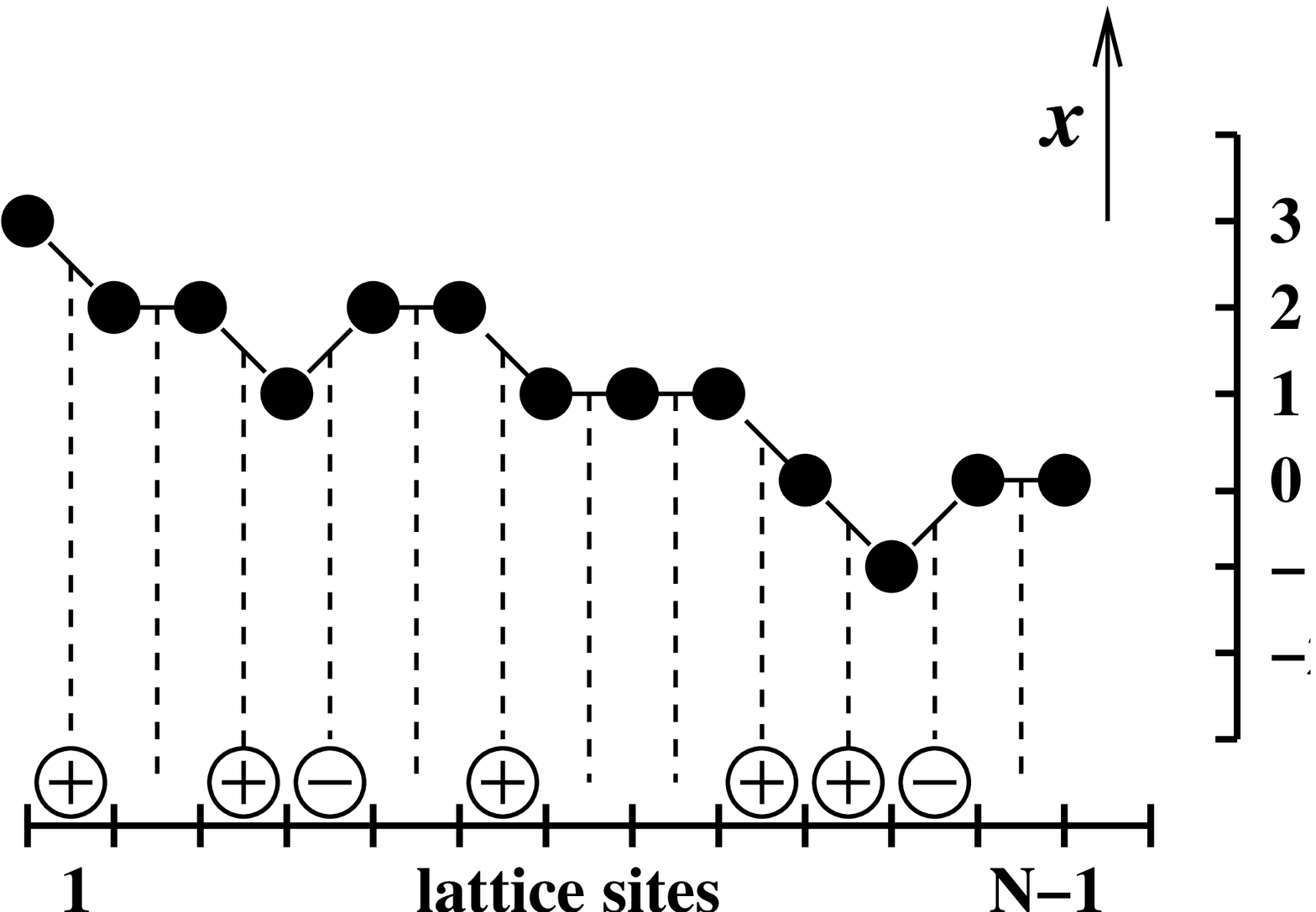}}
\end{picture}
\vskip 3in
 \begin{Large} Fig.2 \end{Large}
\end{center}
\vskip 3in
\end{figure}

\begin{figure}[h]
\begin{center}
\vskip 1.5in
\unitlength 1in
\begin{picture}(4.5,3.5)
\resizebox{4.7in}{4.5in}{\includegraphics{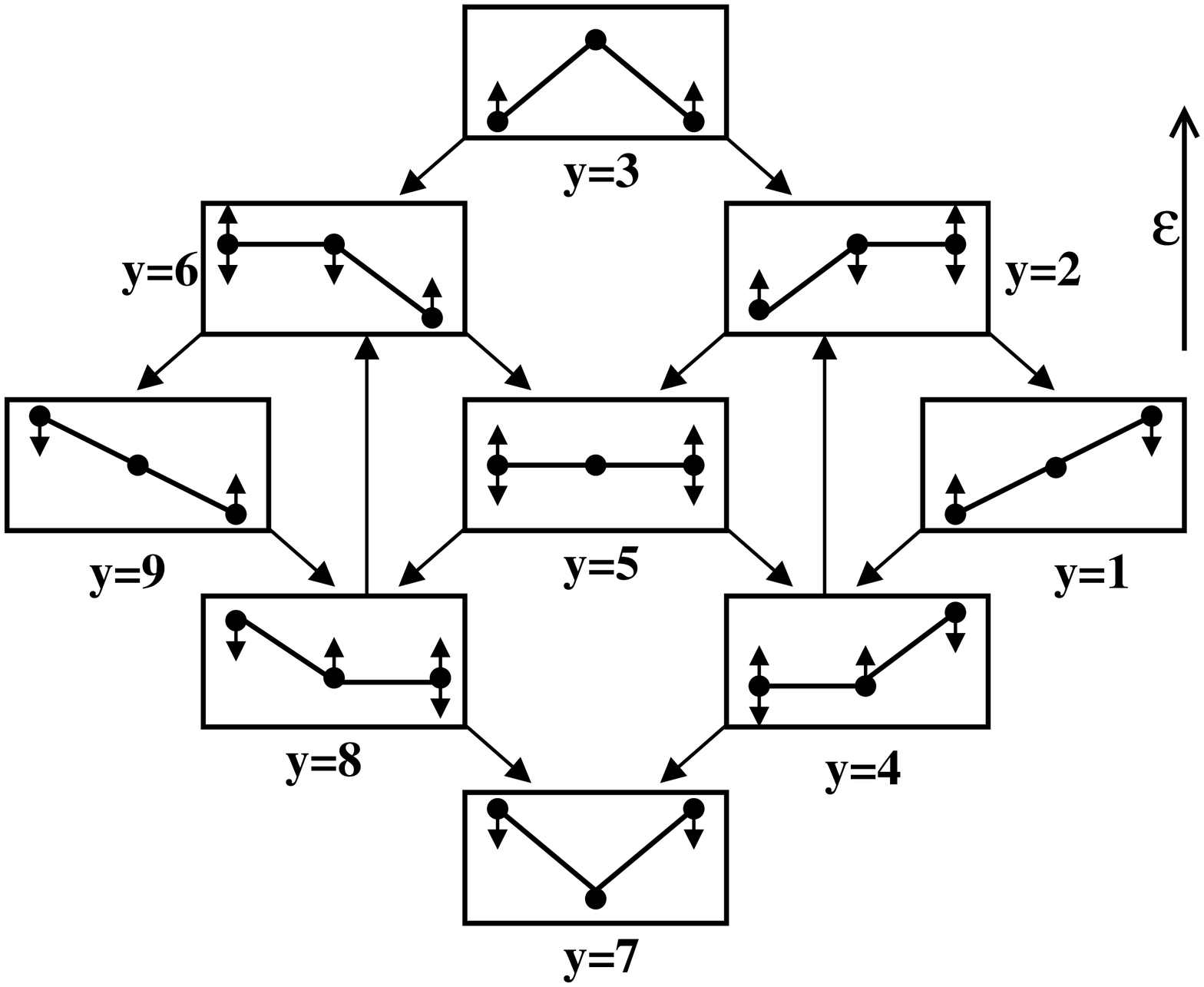}}
\end{picture}
\vskip 3in
 \begin{Large} Fig.3 \end{Large}
\end{center}
\vskip 3in
\end{figure}

\begin{figure}[h]
\begin{center}
\vskip 1.5in
\unitlength 1in
\begin{picture}(4.5,3.5)
\resizebox{4.5in}{3.3in}{\includegraphics{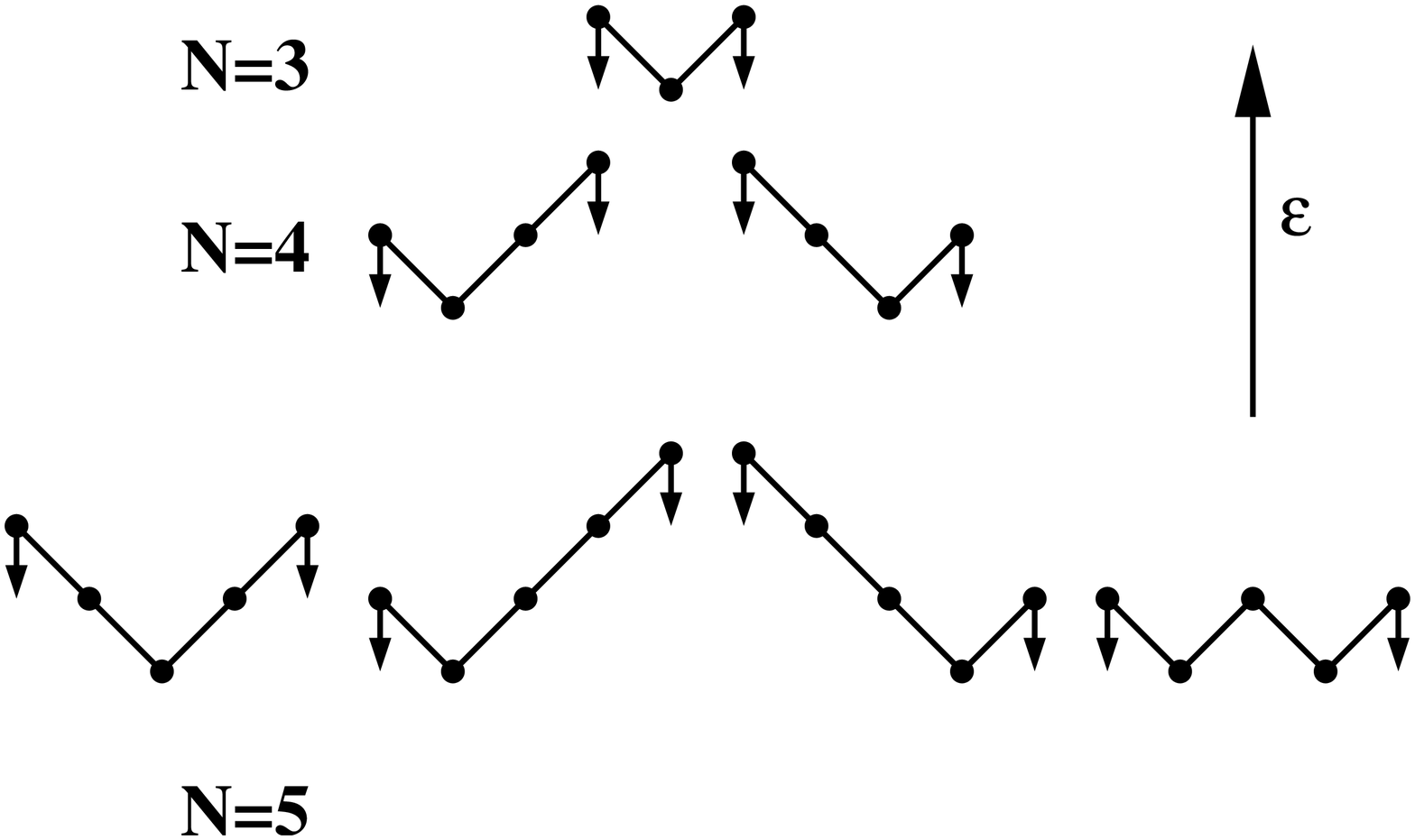}}
\end{picture}
\vskip 3in
 \begin{Large} Fig.4 \end{Large}
\end{center}
\vskip 3in
\end{figure}

\begin{figure}[h]
\begin{center}
\vskip 1.5in
\unitlength 1in
\begin{picture}(4.5,3.5)
\resizebox{4.5in}{3.5in}{\includegraphics{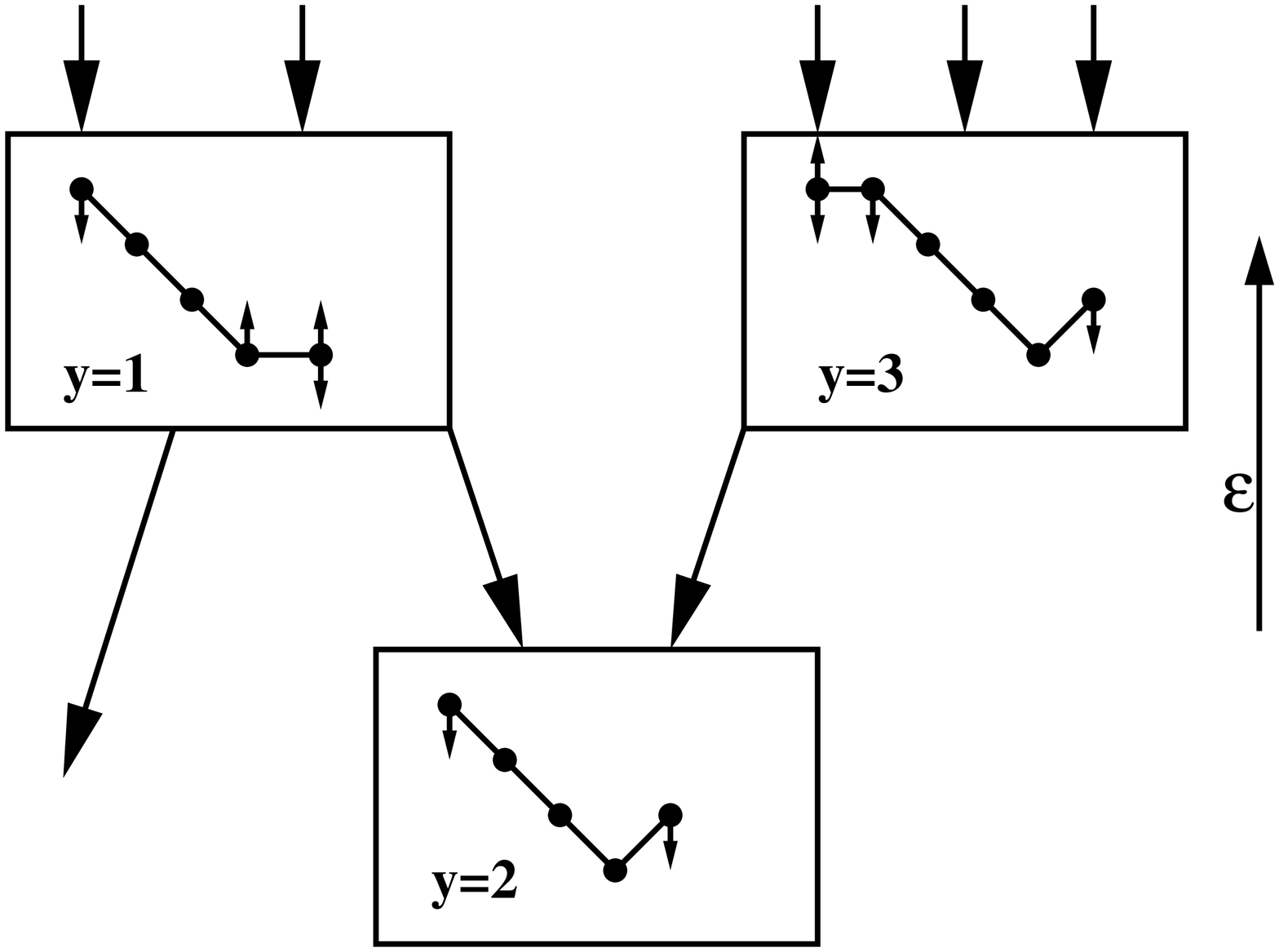}}
\end{picture}
\vskip 3in
 \begin{Large} Fig.5A \end{Large}
\end{center}
\vskip 3in
\end{figure}

\begin{figure}[h]
\begin{center}
\vskip 1.5in
\unitlength 1in
\begin{picture}(4.5,3.5)
\resizebox{4.5in}{3.5in}{\includegraphics{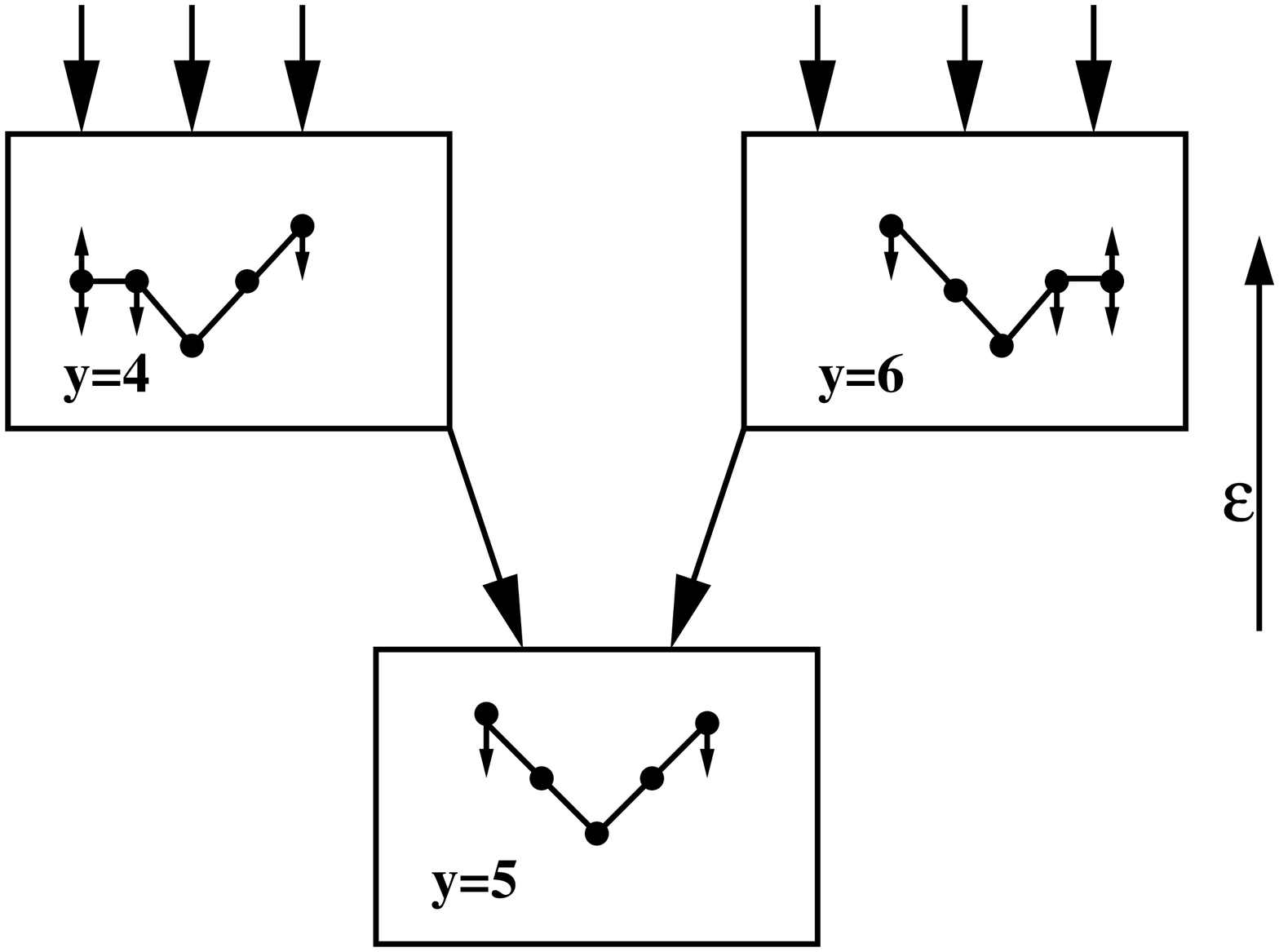}}
\end{picture}
\vskip 3in
 \begin{Large} Fig.5B \end{Large}
\end{center}
\vskip 3in
\end{figure}

\begin{figure}[h]
\begin{center}
\vskip 1.5in
\unitlength 1in
\begin{picture}(6.5,5.5)
\resizebox{6.5in}{5.5in}{\includegraphics{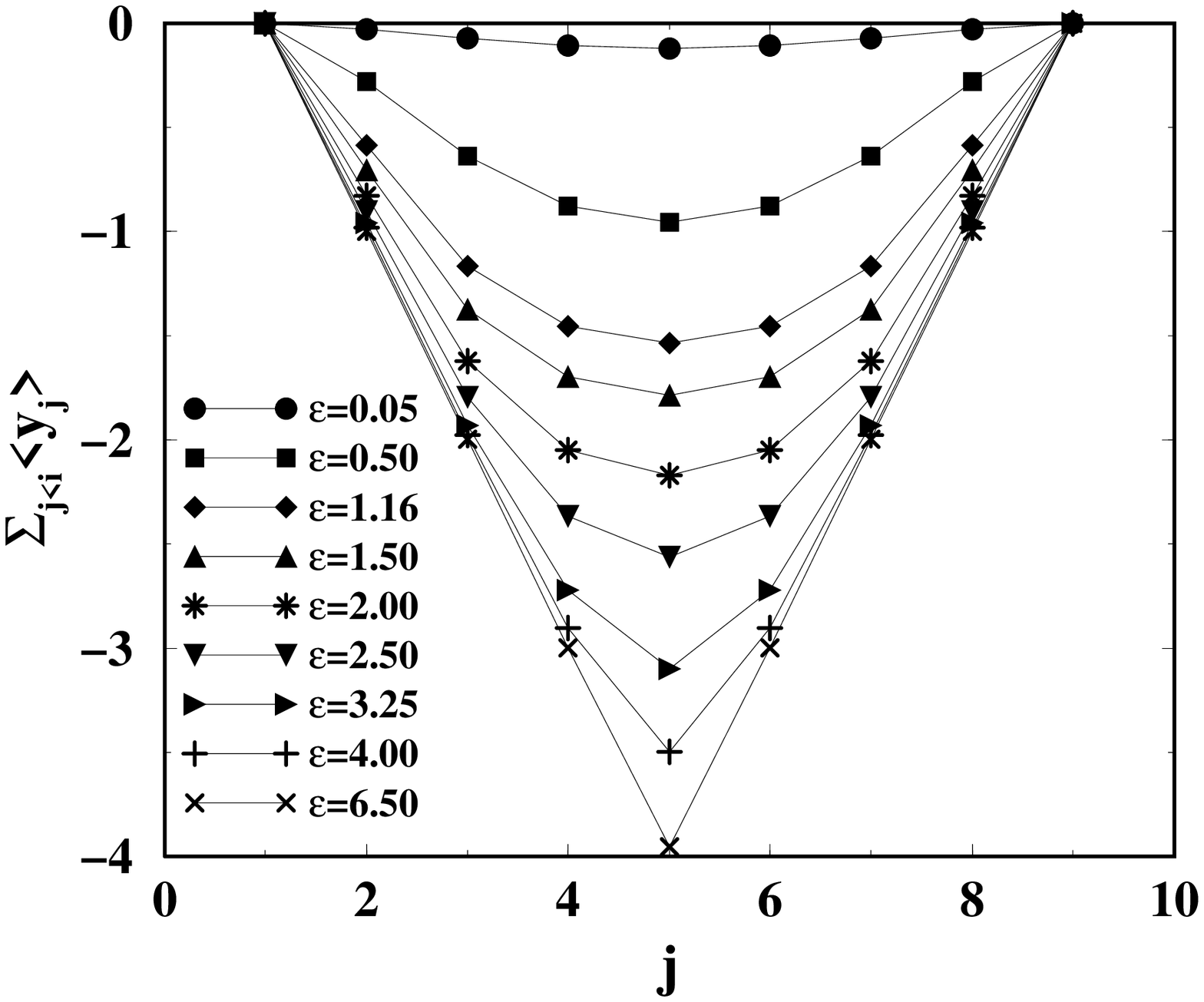}}
\end{picture}
\vskip 1in
\begin{Large} Fig.6 \end{Large}
\end{center}
\vskip 3in
\end{figure}

\begin{figure}[h]
\begin{center}
\vskip 1.5in
\unitlength 1in
\begin{picture}(6.5,5.5)
\resizebox{6.5in}{5.5in}{\includegraphics{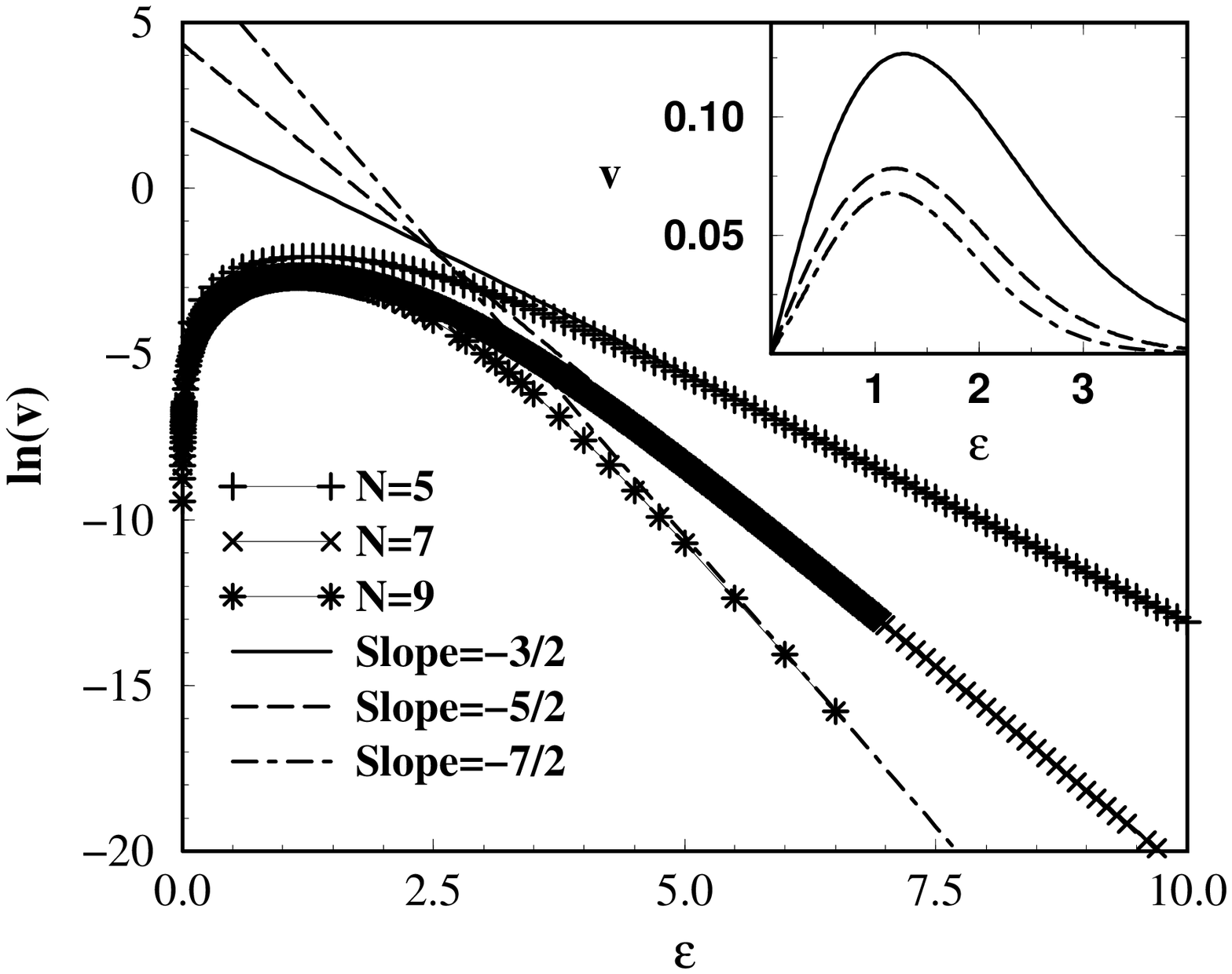}}
\end{picture}
\vskip 1in
\begin{Large} Fig.7 \end{Large}
\end{center}
\vskip 3in
\end{figure}

\begin{figure}[h]
\begin{center}
\vskip 1.5in
\unitlength 1in
\begin{picture}(6.5,5.5)
\resizebox{6.5in}{5.5in}{\includegraphics{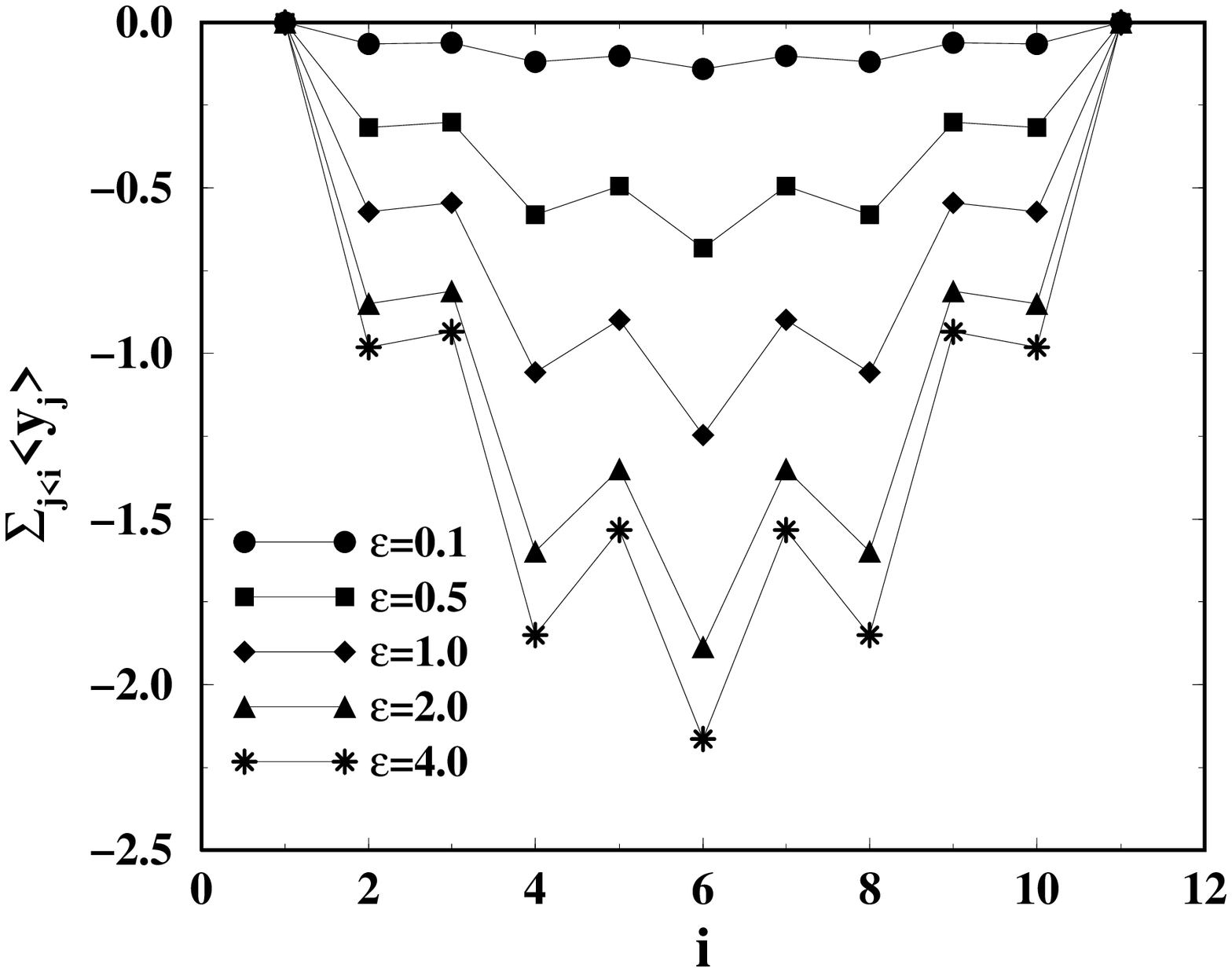}}
\end{picture}
\vskip 1in
\begin{Large} Fig.8 \end{Large}
\end{center}
\vskip 3in
\end{figure}

\begin{figure}[h]
\begin{center}
\vskip 1.5in
\unitlength 1in
\begin{picture}(6.5,5.5)
\resizebox{6.5in}{5.5in}{\includegraphics{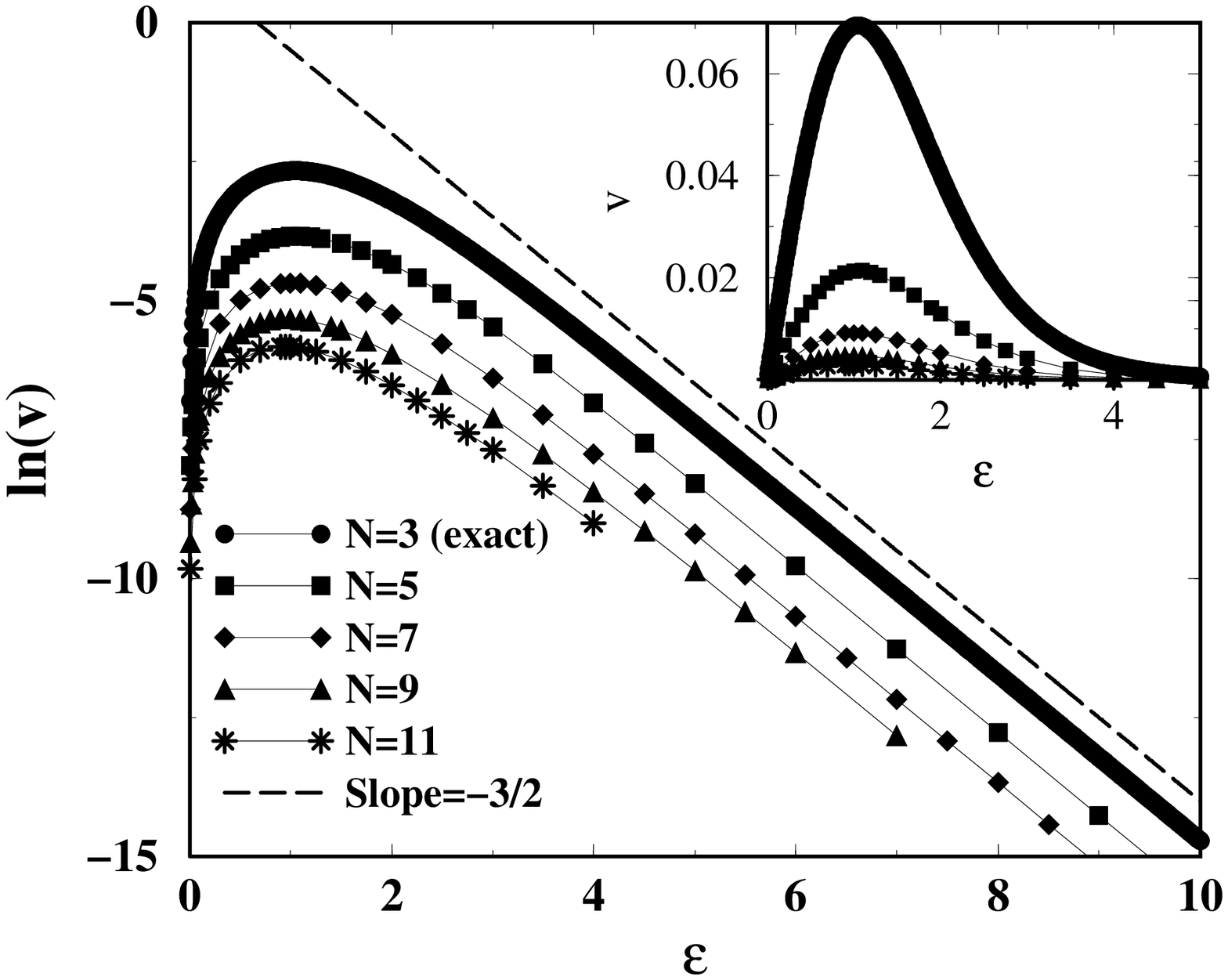}}
\end{picture}
\vskip 1in
\begin{Large} Fig.9 \end{Large}
\end{center}
\vskip 3in
\end{figure}

\begin{figure}[h]
\begin{center}
\vskip 1.5in
\unitlength 1in
\begin{picture}(6.5,5.5)
\resizebox{6.5in}{5.5in}{\includegraphics{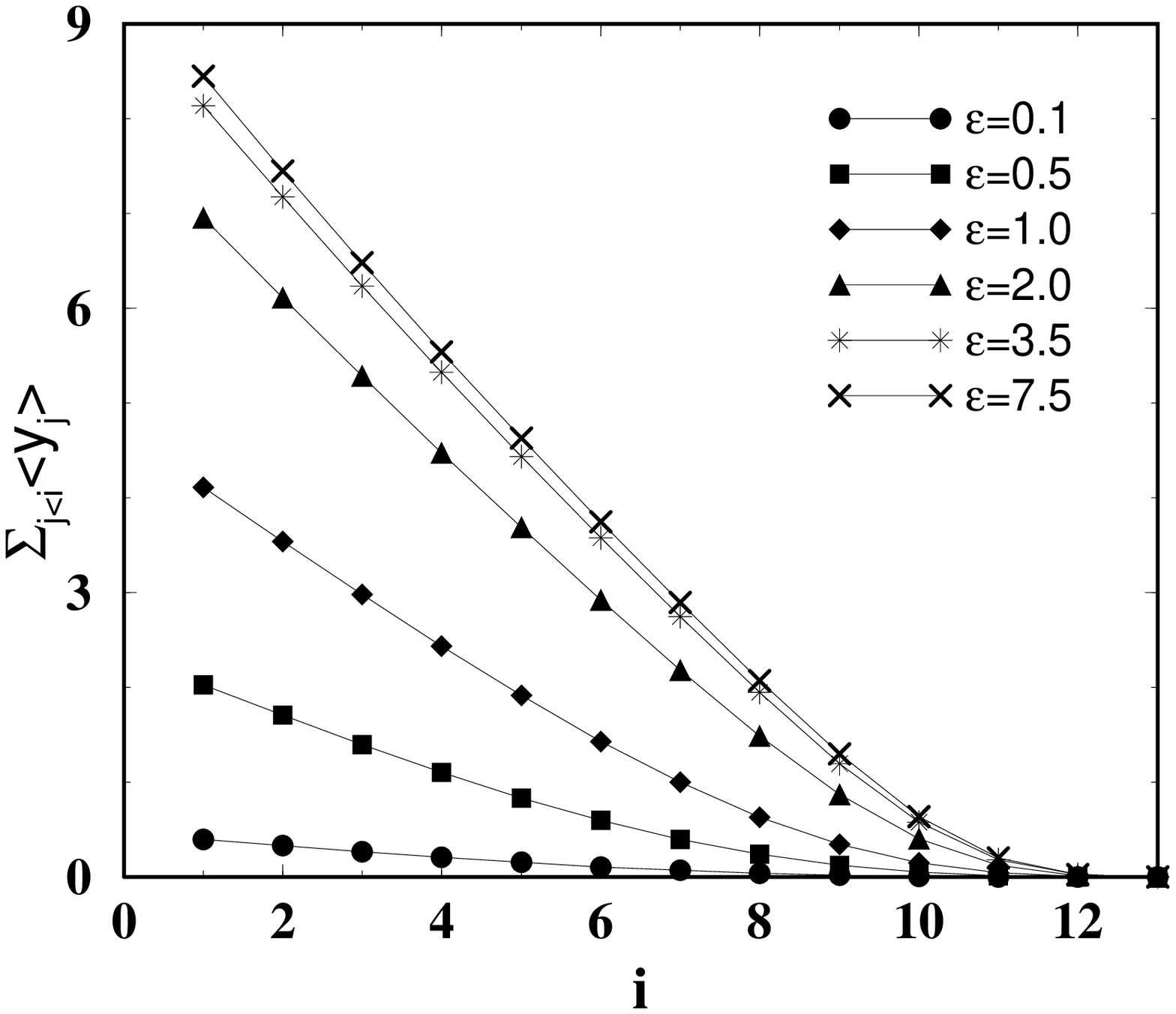}}
\end{picture}
\vskip 1in
\begin{Large} Fig.10 \end{Large}
\end{center}
\vskip 3in
\end{figure}

\begin{figure}[h]
\begin{center}
\vskip 1.5in
\unitlength 1in
\begin{picture}(6.5,5.5)
\resizebox{6.5in}{5.5in}{\includegraphics{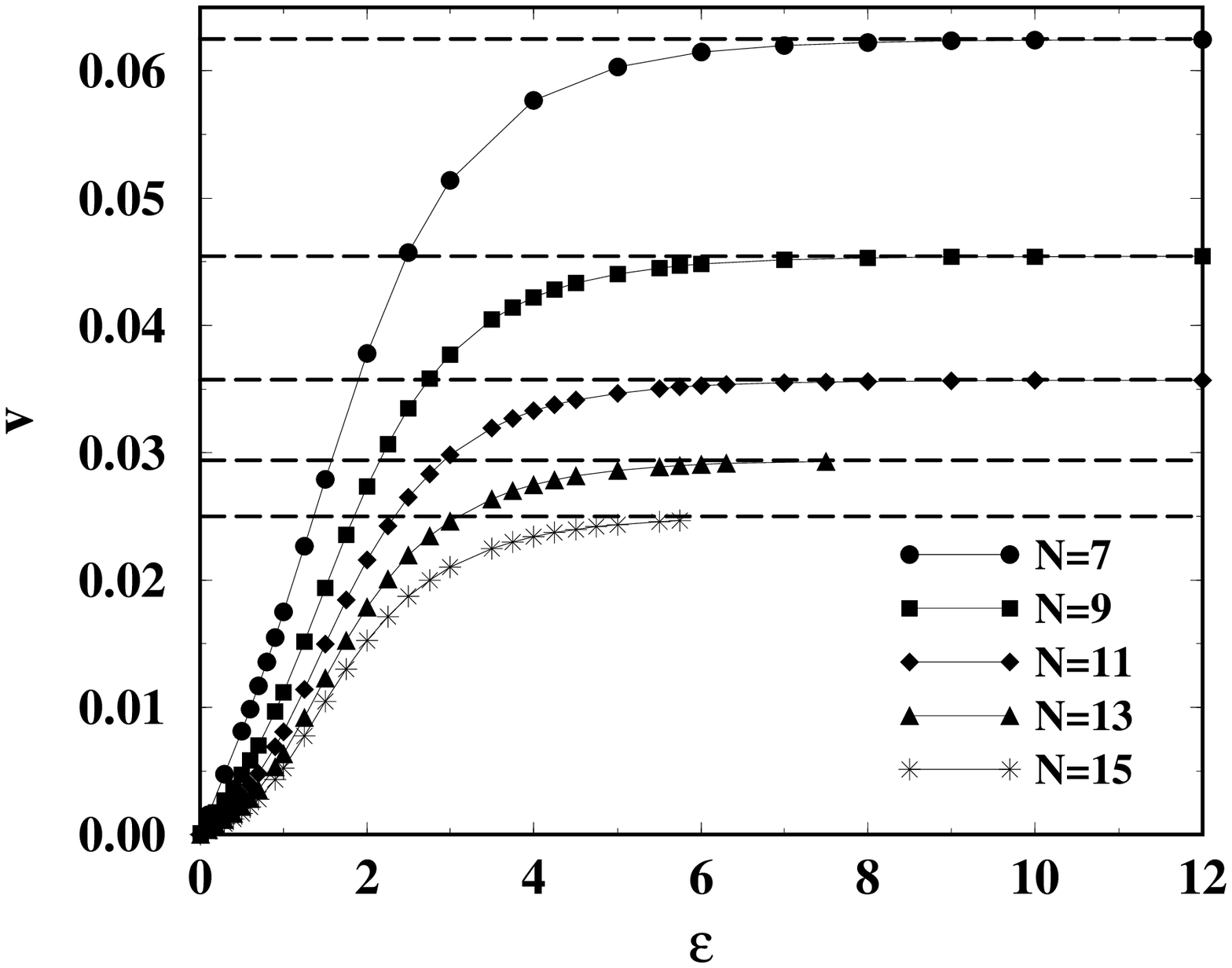}}
\end{picture}
\vskip 1in
\begin{Large} Fig.11 \end{Large}
\end{center}
\vskip 3in
\end{figure}

\begin{figure}[h]
\begin{center}
\vskip 1.5in
\unitlength 1in
\begin{picture}(6.5,5.5)
\resizebox{6.5in}{5.5in}{\includegraphics{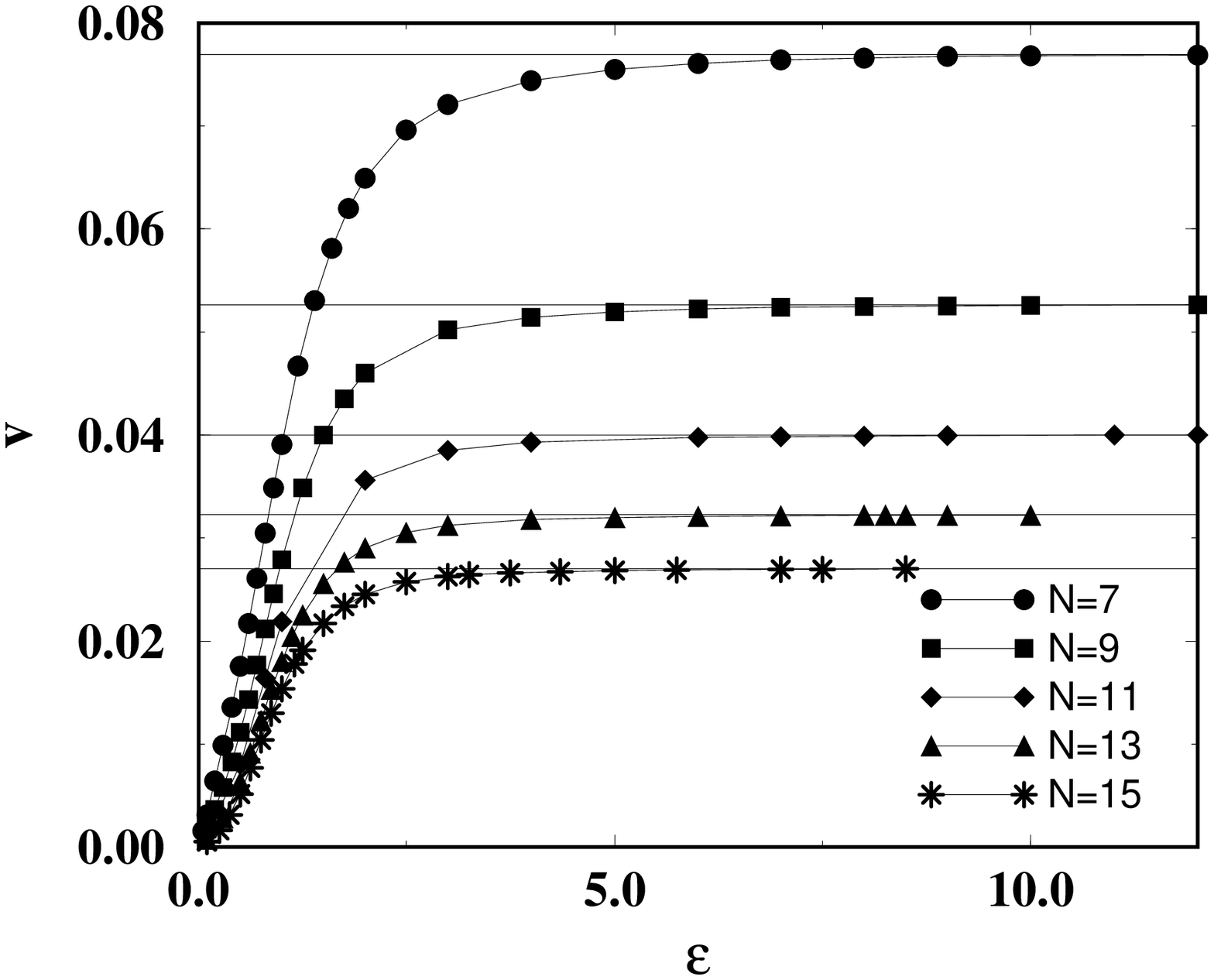}}
\end{picture}
\vskip 1in
\begin{Large} Fig.12 \end{Large}
\end{center}
\vskip 3in
\end{figure}

\newpage

\noindent {\bf Figure Captions:} \\\\

\noindent Fig. 1 a)  Two-dimensional representation of DNA molecule in a 
gel. Crosses represent gel fibers around which the chain is entangled. 
Dots mark the midpoints of segments;\\
b) Repton  representation  of DNA in a gel. A single spatial coordinate $x$ 
is in the direction of an externally applied electric field $\varepsilon$. The arrows 
represent allowed moves.\\
\\
Fig. 2  An illustration of the correspondence between a configuration of the 
repton model, and a configuration in the asymmetric simple exclusion model 
with two species of particles.\\
\\
Fig. 3  Diagrammatic picture for  the  polymer of size $N=3$. Arrows between 
boxes indicate allowed transitions between configurations.\\
\\
Fig. 4 Trap configurations for polyelectrolyte chains of size $N=3$, 4  and 5.\\
\\
Fig. 5 a) Part of the diagram for $N=5$ chain near the non-real trap 
configuration. \\
b) Part of the diagram for $N=5$ chain near the real trap configuration.\\
\\
Fig. 6 The average shape of a polyelectrolyte chain of size $N=9$ for 
different fields. \\
\\
Fig. 7 Plot of the logarithm of a drift velocity as a function of an electric 
field for chains with different length $N$. Inset: the plot $v$ vs. $\varepsilon$
for the same lengths. \\
\\
Fig. 8 The average shape of an alternating chain of size $N=11$. \\
\\
Fig. 9 Plot of the logarithm of a drift velocity as a function of an electric
field for alternating chains with different length $N$. The dependence of a
drift velocity for $N=3$ follows Eq.(\ref{tm3}). Inset: the plot of 
$v$ vs. $\varepsilon$ for the same lengths. \\
\\
Fig. 10 The average shape of a chain of size $N=13$  where all reptons are
neutral except one (left) end repton. \\
\\
Fig. 11 Plot of $v$ vs. $\varepsilon$ for various $N$. The dashed lines
are limiting values of a drift velocity according to Eq.(\ref{velocity.final})
for various $N$ when $k=1$. \\
\\
Fig. 12 Plot of $v$ vs. $\varepsilon$ for various $N$. The dashed lines
are limiting values of a drift velocity according to Eq.(\ref{velocity.final})
for various $N$ when $k=2$. \\
\\

\end{document}